\documentclass[sigconf, authorversion]{acmart}

\usepackage{caption}
\usepackage{subcaption}
\AtBeginDocument{%
  \providecommand\BibTeX{{%
    \normalfont B\kern-0.5em{\scshape i\kern-0.25em b}\kern-0.8em\TeX}}}

\copyrightyear{2022} 
\acmYear{2022} 
\setcopyright{acmlicensed}
\acmConference[ICMI '22]{INTERNATIONAL CONFERENCE ON MULTIMODAL INTERACTION}{November 7--11, 2022}{Bengaluru, India}
\acmBooktitle{INTERNATIONAL CONFERENCE ON MULTIMODAL INTERACTION (ICMI '22), November 7--11, 2022, Bengaluru, India}
\acmPrice{15.00}
\acmDOI{10.1145/3536221.3557034}
\acmISBN{978-1-4503-9390-4/22/11}

\begin{document}

\title[Adaptive User-Centered Multimodal Interaction]{Adaptive User-Centered Multimodal Interaction towards Reliable and Trusted Automotive Interfaces}


\author{Amr Gomaa}
\affiliation{%
  \institution{German Research Center for Artificial Intelligence (DFKI)}
    \city{Saarbr{\"u}cken}
  \country{Germany}
}
\affiliation{%
  \institution{Saarland Informatics Campus}
  \city{Saarbr{\"u}cken}
  \country{Germany}
}

\email{amr.gomaa@dfki.de}

\renewcommand{\shortauthors}{Amr Gomaa}

\begin{abstract}

With the recently increasing capabilities of modern vehicles, novel approaches for interaction emerged that go beyond traditional touch-based and voice command approaches. Therefore, hand gestures, head pose, eye gaze, and speech have been extensively investigated in automotive applications for object selection and referencing. Despite these significant advances, existing approaches mostly employ a one-model-fits-all approach unsuitable for varying user behavior and individual differences. Moreover, current referencing approaches either consider these modalities separately or focus on a stationary situation, whereas the situation in a moving vehicle is highly dynamic and subject to safety-critical constraints. In this paper, I propose a research plan for a user-centered adaptive multimodal fusion approach for referencing external objects from a moving vehicle. The proposed plan aims to provide an open-source framework for user-centered adaptation and personalization using user observations and heuristics, multimodal fusion, clustering, transfer-of-learning for model adaptation, and continuous learning, moving towards trusted human-centered artificial intelligence.

\end{abstract}



\begin{teaserfigure}
     \centering
     \begin{subfigure}{0.4\linewidth}
         \centering
         \includegraphics[width=\textwidth]{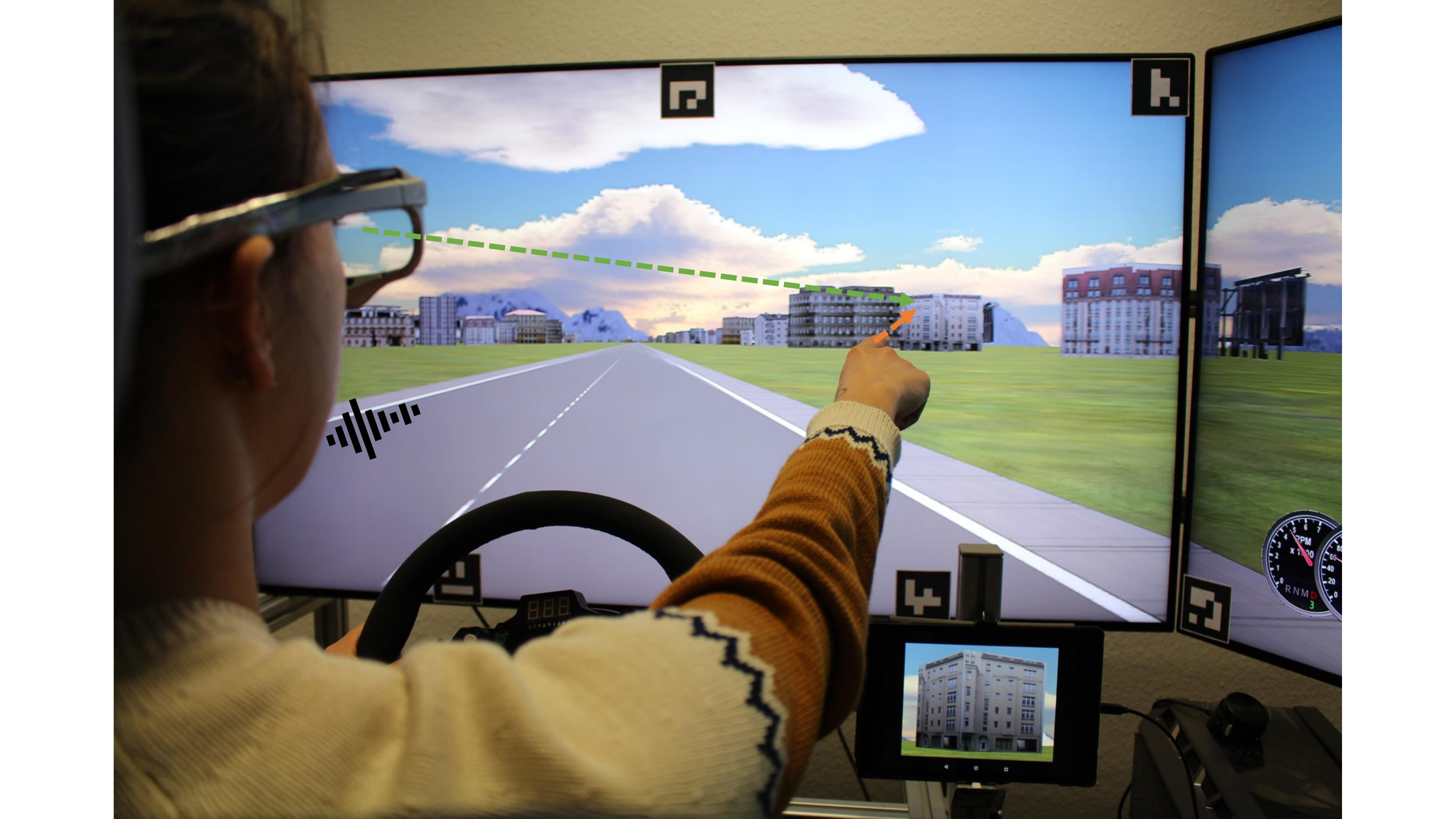}
     \end{subfigure}
     \begin{subfigure}{0.4\linewidth}
         \centering
         \includegraphics[width=\textwidth]{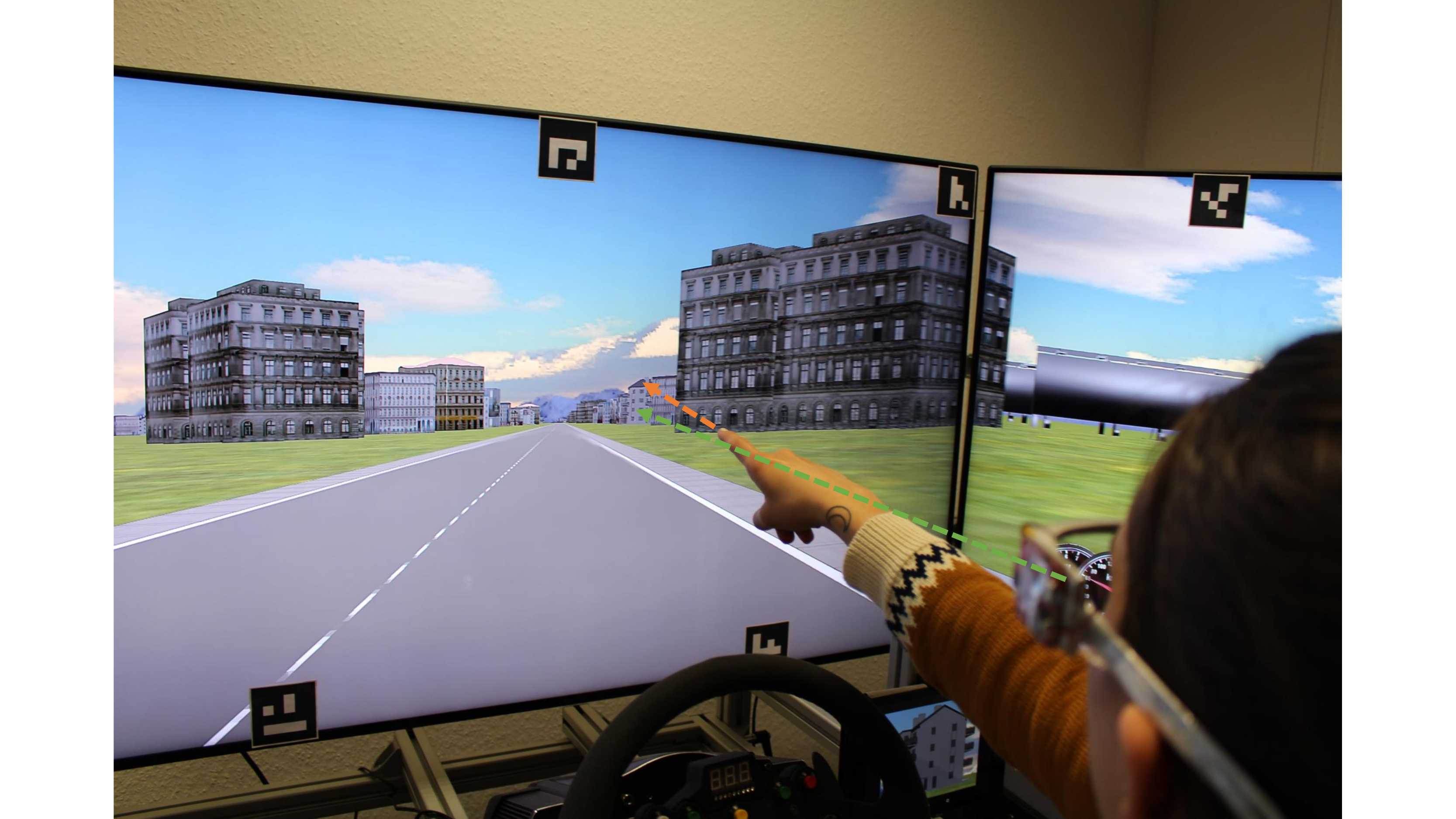}
     \end{subfigure}
     \caption{Multimodal object referencing from a (simulated) moving vehicle.}
     \Description{There are two images showing an example of a driver referencing an object outside the vehicle while driving.}
     \label{fig:visualsearch}
\end{teaserfigure}

\maketitle

\section{Introduction}

Human-centered artificial intelligence (HCAI) is an emerging topic that is rapidly gaining significant interest among both artificial intelligence (AI) and human-computer interaction (HCI) researchers~\cite{xu2019toward,nowak2018assessing,bryson2019society,shneiderman2020human}. The reason for HCAI's recent popularity is that it implicitly and explicitly promotes trust, control, and reliability among users~\cite{shneiderman2020human}. However, implementing HCAI concepts is still an ambiguous and challenging task~\cite{nowak2018assessing}; thus, it is still lacking in recent work~\cite{bryson2019society}. 
As for multimodal fusion, multiple approaches have been implemented utilizing various early and late data fusion techniques~\cite{dong2009advances,karpov2018multimodal}. Specifically for the automotive domain, researchers have focused on multiple aspects of controlling the vehicle and the infotainment system using touch-based approaches and the multimodal combination of several modalities, including hand gestures (e.g., pointing), gaze, and speech~\cite{rumelin2013free,fujimura2013driver,Roider2017,feld2016combine}. Consequently, multiple works incorporate hand gestures and gaze in controlling the infotainment system and various components inside the vehicle~\cite{molchanov2015hand, molchanov2015multi, ohn2014hand, zobl2003real,roider2018see} for their simplicity and naturalness when interacting with a somewhat complicated machine like a modern car. Additionally, recent work has emphasized the importance of the driver's cognitive load when dealing with such interfaces and their effect on driving performance~\cite{Fuller2005TowardsBehaviour,Brown2020Ultrahapticons:Interfaces,Islam2020ALearning,Barua2020TowardsClassification,Solovey2014classifying}, highlighting personalization's significance. However, there is less work focusing on interacting with objects outside the vehicle~\cite{fujimura2013driver,rumelin2013free,Aftab2020}, which is the focus of this work. 

Thus, the proposed research focuses on adaptive and personalized approaches that enhance system performance and promote trust toward a reliable and controllable HCAI. More specifically, I first present initial results on investigating the specific characteristics of each modality and the interaction between them, and highlight individual differences (as published in~\cite{Gomaa2020}). Second, I show a learning-based multimodal fusion approach for referencing using machine learning algorithms and highlight the first step for user-specific adaptation techniques using transfer-of-learning (as published in~\cite{gomaa2021ml}). Finally, I propose a continuous learning approach and an open-source framework for model adaptation based on the driver's implicit and explicit feedback in the object referencing task.

\section{Background and related work}

Referencing and selecting objects have been studied in multiple domains~\cite{nickel20043d,kehl2004real,ji2002real,rumelin2013free}; however, this work focus specifically on the automotive domain. First, single modality object referencing is highlighted for pointing and gaze (most common modalities for object selection), and then multimodal referencing is described. Finally, a brief overview of the adaptation approaches in multimodal interaction is discussed.

\paragraph{Object Referencing using Pointing Gestures}
In-vehicle and outside-the-vehicle object selection using hand pointing gestures was studied with constrained-hand pointing~\cite{fujimura2013driver} and free-hand pointing~\cite{rumelin2013free}. 
Fujimura et al.~\cite{fujimura2013driver} investigated constrained-hand pointing in a simulation environment; they suggested using constrained-hand pointing instead of free-hand pointing to decrease the risk of the driver's hand being taken off the wheel. This came with many limitations on the accurate tracking of the pointing vector, and they did not report exact tracking accuracy figures due to unrealistic approximations.
Consequently, Rümelin et al.~\cite{rumelin2013free} investigated free-hand pointing using two approaches: a lab study with a stationary car and street scenes presented on multiple projectors and a field study using the Wizard of Oz technique to collect qualitative feedback. They successfully tracked the driver's pointing gestures in the lab using a low-resolution depth camera, whereas the same setup did not work for the field study. However, they could draw qualitative conclusions about driver behavior when performing the dual task of driving and referencing objects. They identified several aspects regarding the driver's gazing focus when performing the pointing gesture, although they were not explicitly investigating the multimodal interaction. They defined three gazing phases, named \textit{information glance}, \textit{pointing positioning}, and \textit{control glance}. These gazing phases inspire the first step of this proposed research work: analyzing and modeling the driver's behavior during the multimodal referencing task~\cite{Gomaa2020}.


\paragraph{Object Referencing using Head Pose and Eye Gaze}

Several studies were conducted to monitor a driver's activity using head pose and eye gaze tracking~\cite{ji2002real, ohn2014head, vicente2015driver, vasli2016driver, vora2017generalizing}; however, few focused on the object selection task~\cite{poitschke2011gaze, kang2015you}.
Kang et al.~\cite{kang2015you} utilized head pose and eye gaze for referencing objects outside the vehicle using a depth camera in a field study. Similar to referencing using pointing gestures, they considered only the horizontal angles for the referencing task. The estimated referencing angle was the summation of the car orientation, head pose, and eye gaze angle. Due to their camera position (behind the steering wheel), gaze detection suffered greatly. Thus, the advantage of a multimodal fusion approach is quite clear here, even in the simplest form of modality switching when the other modality data is unavailable.
Similarly, Poitschke et al.~\cite{poitschke2011gaze} studied in-vehicle object selection and compared it to traditional touchscreen interaction. They utilized a button attached to the steering wheel to determine the onset of the selection task. They showed a significant increase in the selection speed with their approach compared to the touchscreen one. However, they also showed that the driver's cognitive load significantly increased during this shorter period. This load could be alleviated with a more natural form of interaction using adapted models (e.g., complementing gaze with speech or pointing instead of the button press, or explicitly switching off the tracking of modalities causing high mental workload) as proposed in this research work.

\begin{figure*}[t]
	\begin{center}
		\includegraphics[width=0.8\linewidth]{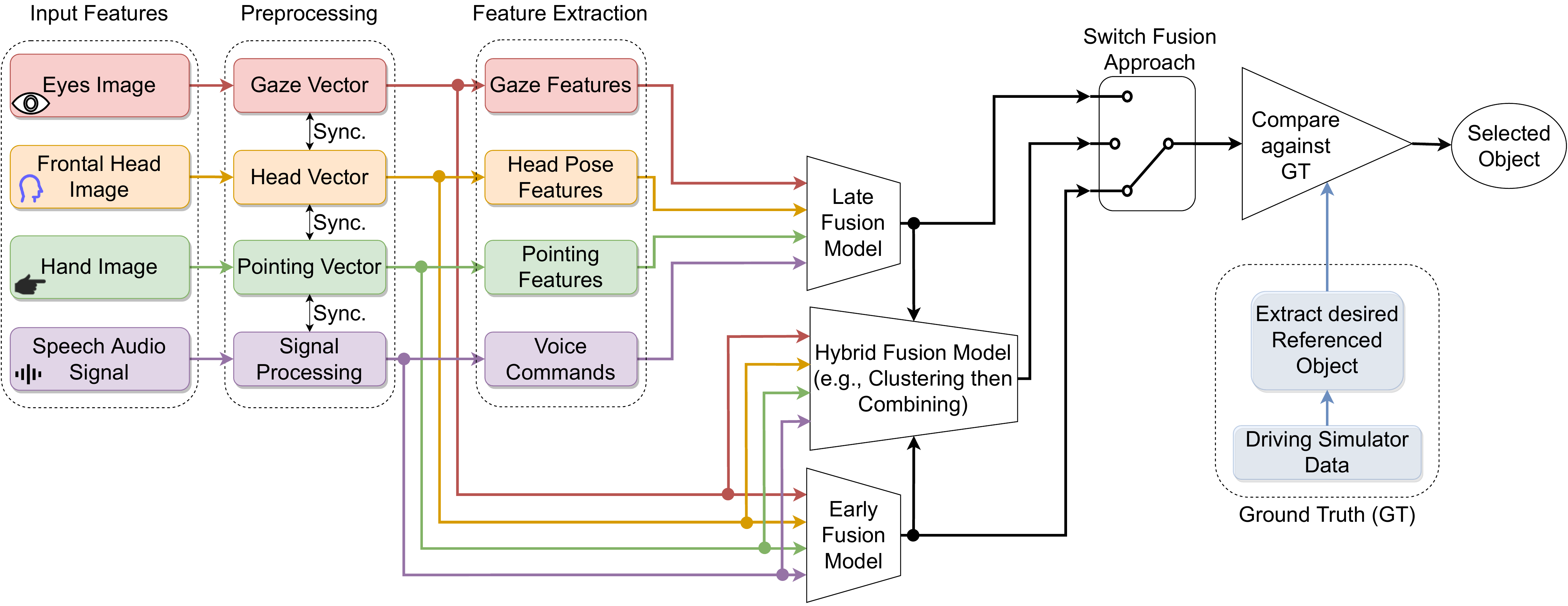}
	\end{center}
	\caption{The proposed generic system architecture for different fusion approaches.}
	\Description{The proposed system architecture. The architecture takes all the modalities as inputs (e.g., eyes images, hand images and speech audio) then transform them to a corresponding signal (e.g., gaze and pointing vectors or different audio signals) in a synchronized preprocessing step. Next, features are extracted to be inputted to the learning algorithm and the output is compared to the ground truth to select the referenced object.}
	\label{fig:implementationarchitecture}
\end{figure*}

\paragraph{Multimodal Object Referencing}

Researchers attempted multimodal fusion approaches for in-vehicle object selection in multiple works~\cite{roider2018see,Aftab2020,sezgin2009multimodal}. However, in-vehicle object referencing approaches do not generalize directly to outside-the-vehicle referencing, as the object's environment is static, limited, and in close proximity. Consequently, Moniri et al.~\cite{Moniri2012a} studied the single task of outside-the-vehicle referencing from the passenger seat using pointing, head pose, and eye gaze. Similarly, Aftab et al.~\cite{aftab2021multimodal} combined these modalities using a neural network-based late fusion approach for referencing from a stationary vehicle. While both approaches showed great promise, they still considered a stationary single-task situation. This work proposes a dual-task scenario of outside-the-vehicle object referencing while driving. A machine learning approach versus a neural network one and early versus late fusion techniques are compared in~\cite{gomaa2021ml}, which will be expanded with more focus on the adaptation and continuous learning aspects.

\paragraph{Adaptive Multimodal Interaction}

Despite the previously discussed significant advances in the multimodal referencing task, an adaptive user-centered approach is lacking. Thus, an important goal and factor of the proposed research work is adaptation and user-specific personalization. Adaptive multimodal interaction combining speech, hand gestures, and gaze has been a topic of interest for the research community for the last 20 years in multiple domains, including robotics and automotive applications~\cite{rogers2000adaptive,hassel2005adaptation,janarthanam2014adaptive,manawadu2017multimodal,zhang2015costs,neverova2015moddrop,gnjatovic2012adaptive}. Therefore, it is worth investigating the incorporation of adaptation techniques to the task of object referencing.


\section{Research Questions and Hypotheses}

In line with the previous motivation and related work, the proposed research aims to develop a framework for an adaptive multimodal fusion approach for outside-the-vehicle object referencing from a moving vehicle. The driver would interact seamlessly with the vehicle's system to reference an outer object using pointing gestures, head pose, gaze, and/or speech. Model adaptation based on the driver's individual behavior would enhance object recognition and increase system reliability and user trust. Several research questions are developed to address the previously mentioned research gap and overcome its inherited challenges.


\begin{itemize}
    \item \textbf{User Observations:} How much does the driver's object referencing behavior differ in performance (e.g., timing, precision, and modalities synchronization)? Furthermore, how can it be quantified to separate drivers into meaningful clusters based on individual behavioral variances?
    
    \item \textbf{Multimodal Fusion:} What system and interface design aspects can be utilized for optimum utilization of the given modalities in terms of fusion techniques, temporal dependencies, and learning models to detect the referenced object?
    
    \item \textbf{Adaptation and Personalized Models:} How can the system adapt to user-specific task performance to optimize the referenced object detection? How much does the referencing affect the driver's cognitive load regarding driving quality, and how can the system adjust? What adaptation techniques (e.g., clustering or transfer-of-learning) can be utilized to generate user-centered personalized learning models?

    \item \textbf{Continuous Learning:} How can the system be designed to continuously gather feedback from the user (both implicitly and explicitly) to guarantee constant development and enhancement of the underlying algorithms? How would that affect the system's reliability and user trust?

\end{itemize}

\section{Research Plan and Methodology}

The proposed research plan and method follow the previously mentioned research questions regarding time and work plan. Thus, a multistage approach is planned to reach the adaptive system with continuous driver feedback as follows.

\subsection{User Observations}

The proposed plan's first stage is to understand drivers' behavior variances when performing the multimodal referencing task. For example, some drivers might be pointing without looking at the object (relying on their peripheral view only), while others might stare at it for a prolonged time or take multiple glances at it. Thus, a behavioral model is constructed to systematically categorize and quantify drivers' actions and predict future ones. Multiple dependent variables (DV) are defined to analyze and assess the referencing task for each modality separately; these are categorized into two categories: \textit{performance-related DV} and \textit{timing-related DV}. Then, heuristic approaches are suggested to combine the modalities utilizing their dependencies and individual task performance. Initial results are published in~\cite{Gomaa2020}.

\subsection{Multimodal Fusion}

\begin{figure*}[t]
	\begin{center}
		\includegraphics[width=0.75\linewidth]{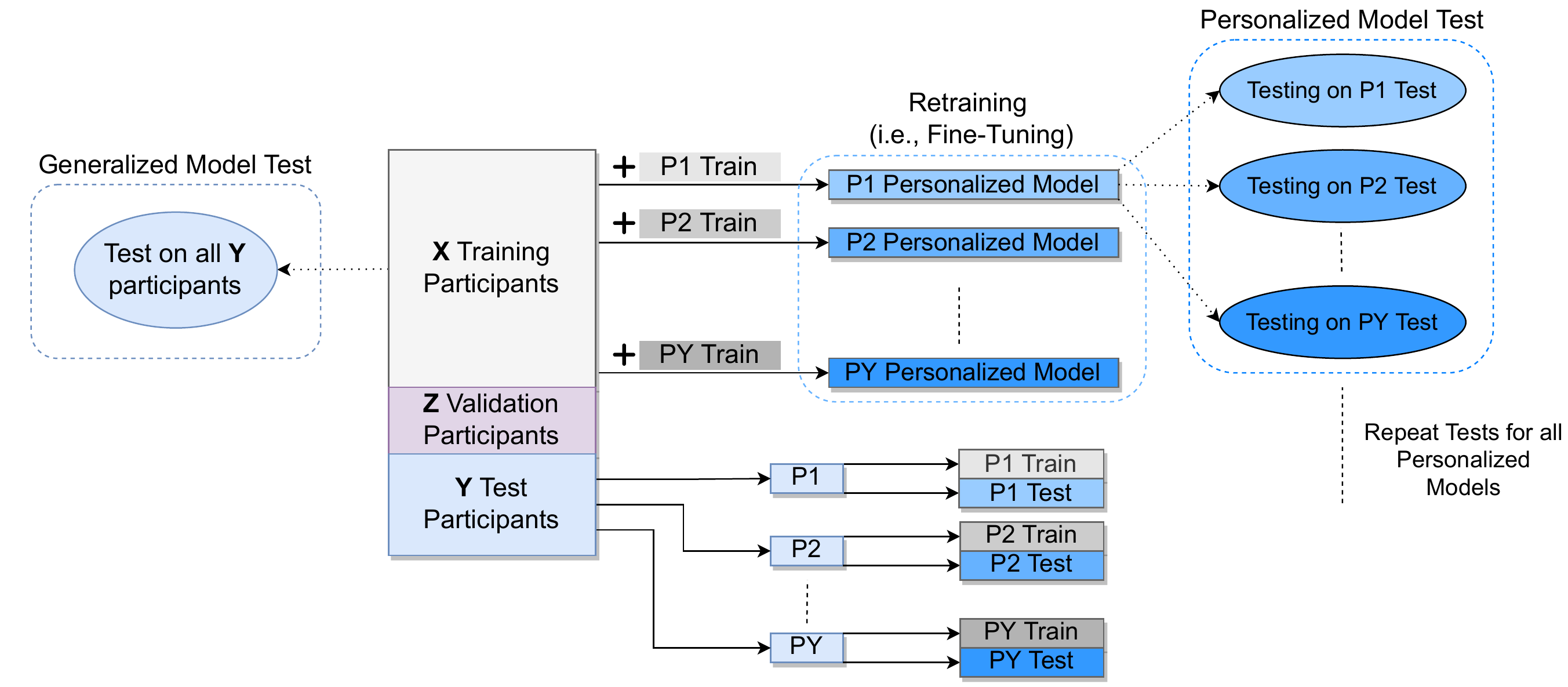}
	\end{center}
	\caption{The proposed approach for model adaptation for generating personalized models through transfer-of-learning.}
	\Description{The proposed approach for model adaptation. The image shows a data split to X training set, Z validation set, and Y test set. It shows the conventional learning approach (hereafter called Generalized Model) of training and validating on X and Z respectively, then testing on Y. The image also shows the personalized approach of further splitting the Y test data participant-wise to additional sub-train and sub-test sets and fine-tuning the training for this specific participant. Finally, the personalized models are tested on each of the participant sub-test data separately for comparison.}
	\label{fig:personalization}
\end{figure*}

The second stage is creating an end-to-end multimodal fusion framework for the object referencing task. This work aims to exhaustively investigate the interaction between the given modalities in terms of performance, timing, user behavior, and fusion techniques.~\autoref{fig:implementationarchitecture} highlights the proposed architecture for combining the given modality for the referencing task. While the well-established, widely used data fusion approaches, such as late and early fusion approaches, are utilized here, more novel and empirical hybrid approaches are also considered that combine heuristics with learning-based data fusion to achieve optimum performance. Additionally, there exists a timing dependency (e.g., modalities' relative onset) between the modalities that the system can exploit. Thus, the time frames can be analyzed separately with no connection, or a pattern could be learned from intra- (within the modality) and inter- (among the modalities) dependencies. Lastly, multiple possible machine learning solutions (depending on the previous factors) are utilized and planned for this work. While preliminary results are published in~\cite{gomaa2021ml}, several expansions are planned regarding learning models, fusion approaches, and timing analysis.

\subsection{Adaptation and Personalized Models}

As mentioned earlier, adaptation and personalized models play an essential role in enhancing the system performance and adjusting to the driver's individual needs. Adaptation is an inherently continuous paradigm; thus, it is considered an ongoing process along the user observations and the multimodal fusion stages in drivers' categorization (i.e., clustering) and hybrid fusion approaches, respectively. While adaptation, in the previous context, is one alternative to the one-model-fits-all approach, it still groups users in a particular model, constituting a many-models-fits-all approach. However, personalized models attempt to have one exact model for each user that is adjusted and adapted to user-specific behavior.~\autoref{fig:personalization} shows an approach to achieving these personalized models through the transfer-of-learning paradigm. The dataset is initially split into training, validation, and test sets as in traditional learning approaches. The model is trained on $\textbf{X}$ participants' data while the hyperparameters are chosen and validated on $\textbf{Z}$ participants' data, and the final model is tested on $\textbf{Y}$ participants' data. On the other hand, for the adaptation approach, each participant's data from the $\textbf{Y}$ test set is further split (e.g., equally) into sub-train and sub-test sets where the model is retrained and fine-tuned on the user-specific training data to produce personalized model weights that are optimized for this user. To assess the effect of this approach, the personalized model is tested on the same participant sub-test data and compared against other participants' sub-test data. Preliminary results for this approach are published in~\cite{gomaa2021ml}.

\subsection{Continuous Learning}

While the previous approaches optimize the system performance in object detection based on current individual behavior, this behavior might change over time due to situational, emotional, or mental load variations and learning effects. Thus, a continuous learning approach is considered where the user can give feedback to the system implicitly (e.g., via dissatisfied looks or grunting as visual or auditory cues) or explicitly (e.g., repeating the given voice command). In this work, an end-to-end learning-based approach is proposed where the model (in terms of weights) and the learning technique (in terms of modalities and timing) would be adapted over time. To achieve this goal, the study and data collection phase should include different situational and mental state variations for internal and external validity. For example, while the referenced objects have some variations in their shape and surrounding environment as currently implemented in this work so far~\cite{Gomaa2020,gomaa2021ml}, other parameters could be varied, such as traffic flow (e.g., a road accident), system response (e.g., intended delayed response), noise level (e.g., talking passengers), and driver's state-of-mind (e.g., driving while angry) as currently planned for this work. Finally, situation-adapting learning techniques are investigated in parallel, such as graph classification and node selection (e.g., Relational Graph Neural Networks~\cite{jing2020relational}), learning from the driver's behavior (e.g., Efficient Learning from Demonstrations~\cite{li2022efficient}), and learning from the driver's feedback (e.g., Implicit Human Feedback Learner~\cite{cui2020empathic}).

\section{Preliminary Results and Planned Work}

\begin{figure}[b]
     \centering
     \begin{subfigure}{\linewidth}
         \centering
         \includegraphics[width=0.85\linewidth]{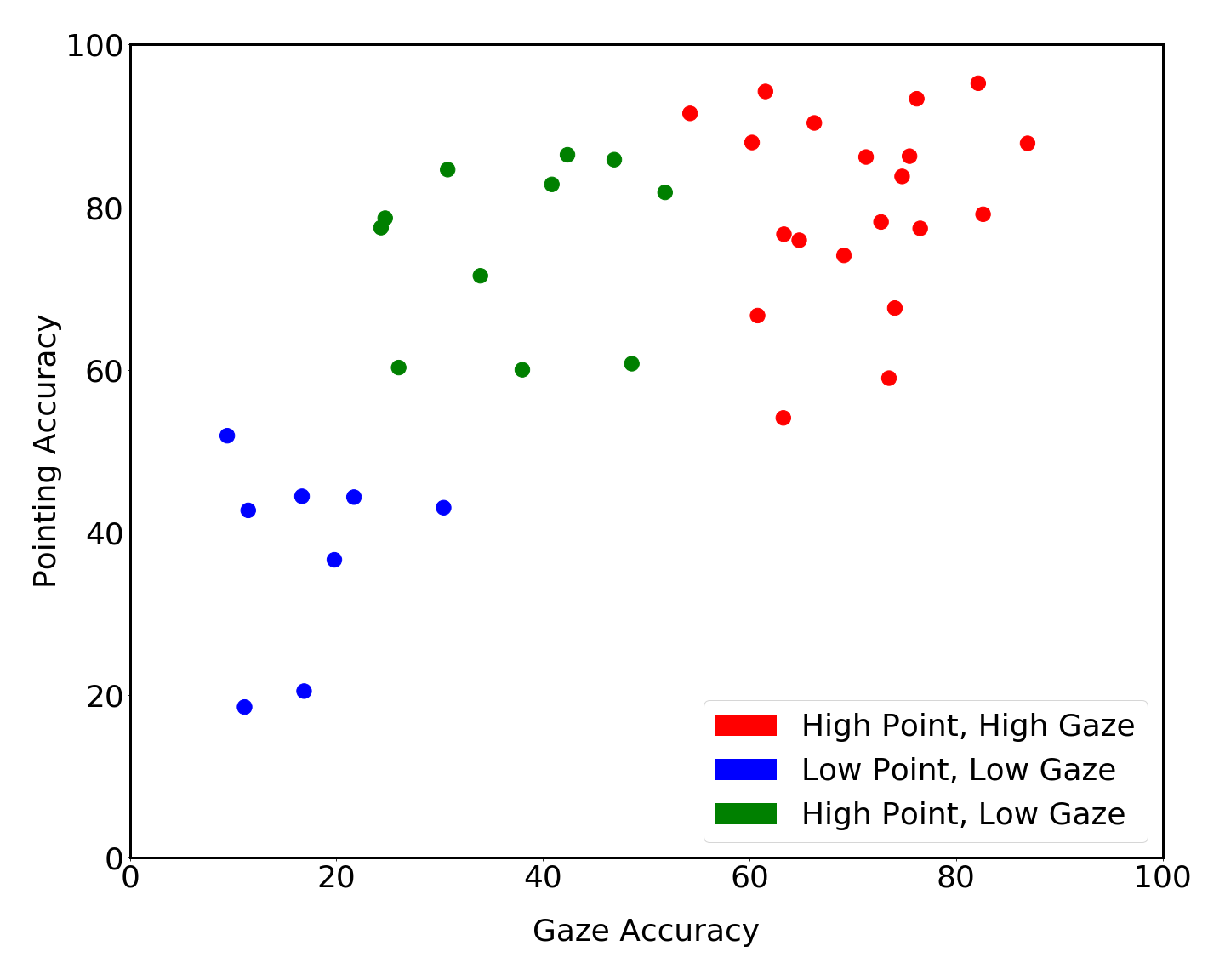}
         \caption{Clustering drivers' pointing and gaze behavior based on the system's perceived performance (i.e., referencing accuracy).}
         \Description{The graph shows a scatter plot of gaze accuracy (on the x-axis) and pointing accuracy (on the y-axis). It shows that in can be clustered into three clusters based on the perceived performance. The clusters are: high pointing and high gazing, low pointing and low gazing, and high pointing and low gazing. The three clusters are almost equally weighted.}
         \label{fig:clustering}
     \end{subfigure}
    \hspace{0.8cm}
     \begin{subfigure}{\linewidth}
         \centering
         \includegraphics[width=0.85\linewidth]{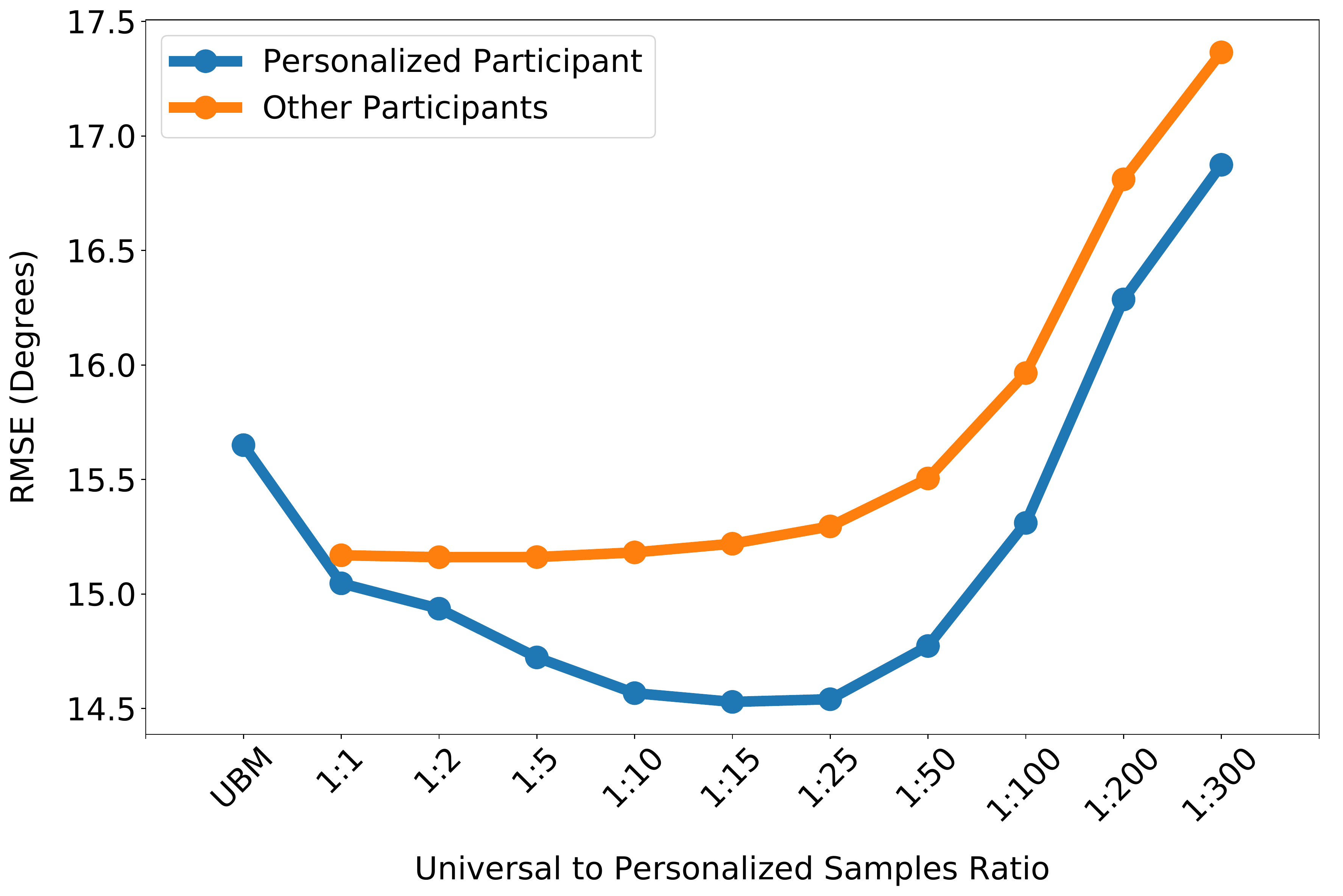}
         \caption{RMSE results comparing different personalized models among themselves (based on the sample weight of the user-specific data) and against the UBM (i.e., generalized model).}
         \Description{The graph shows two line plots of the RMSE results of the participant wise sub-tests compared at difference sampling ratio values. The x-axis is the sampling ratio spanning from ``1:1'' until ``1:300'' while the y-axis is the RMSE value in degrees. One line plot is the personalized participant test results while the other line plot is the average of the test results for the other participants. It shows the gap between the lines increase while increasing the sampling ratio, however, until it starts over fitting the RMSE results get worse for both line plots. }
         \label{fig:personalizationsampleweight}
     \end{subfigure}
     \caption{Examples of different adaptation and personalization approaches from the initial results published in~\cite{Gomaa2020,gomaa2021ml}.}
     \label{fig:resultsexamples}
\end{figure}

Several studies have already been conducted to achieve this work's goals, with initial results published in~\cite{Gomaa2020,gomaa2021ml}. Since the core focus of this work is on adaptation and user-specific personalization,~\autoref{fig:resultsexamples} shows examples of the preliminary results focusing on the adaptation aspect. Specifically,~\autoref{fig:clustering} shows how drivers' referencing actions could be clustered based on pointing and gaze modality performance separately; then, each cluster is trained independently. Thus, each cluster model-weights would be adapted to the cluster pointing- and gaze-specific accuracy. This resembles the hybrid fusion approach discussed earlier. Similarly,~\autoref{fig:personalizationsampleweight} highlights the results of the previously discussed transfer-of-learning personalization approach. It compares the personalized model sub-test data against the average of the other non-personalized sub-test data using the Root Mean Square Error (RMSE) metric. The figure also highlights further enhancement of this personalization approach; it was noticed that adding the sub-train data of the personalized participant to the existing generalized model (also called Universal Background Model (UBM)) data with a 1:1 ratio is not the optimum solution due to its insignificant contribution size. Thus, the personalized participant sub-train data was emphasized (e.g., by repeating the data multiple times), and its ratio increased for the $\textbf{X}$ training data with a ratio of 1:2, 1:5, and so on until the optimum sample weight could be determined. 
These previous results are a sample of possible adaptation strategies for object referencing using pointing and gaze; however, further studies are planned that would incorporate other modalities such as head pose (independent from gaze) and speech as well as other adaptation techniques. Consequently, some of the expected results include (1) explicit comparison between co-dependent modalities (e.g., gaze and head pose) in terms of their importance to the fusion model, (2) enhanced model performance with more powerful deep learning algorithms (e.g., attention-based neural networks~\cite{vaswani2017attention}), and (3) better user-specific object referencing accuracy when utilizing the situation-aware adaptation methods discussed earlier. 

\section{Current and Expected Contributions}

While the core objective of this work is to reach an adaptive, personalized approach for referencing objects while driving, its contributions have several folds, as follows.

\begin{enumerate}
    \item \textbf{Empirical Hypotheses:} The literature review has provided multiple, albeit unverified, hypotheses on the multimodal interaction in general and the referencing task in particular. This work systematically investigates and verifies such hypotheses statistically and qualitatively and explores new hypotheses to model users' behavior.
    \item \textbf{Design Insights and Guidelines:} This work provides strategies and approaches for the referencing task's generalized and user-specific multimodal interaction. These strategies could also be utilized in other multimodal interaction tasks that involve similar modalities with minor alterations.  
    \item \textbf{Methodological Contribution:} An open-source framework for adaptive and personalized models is tailored for user-specific behavior. Furthermore, this framework would focus on continuous learning approaches invariant to users' varied behavior. The first milestone of this framework is already published and publicly available~\cite{gomaa2021ml}.
    \item \textbf{Dataset Contribution:} Due to the importance of open science and results' reproducibility, the data sets for this work are published (in anonymous form) for inspection, verification, and future work continuity.
    
\end{enumerate}

    Finally, multiple aspects could be considered when designing user-specific interfaces that go beyond what is proposed in this work. However, this work investigates many of these aspects in learning model adaptation, modality exploitation, and system engineering and highlights important factors for future work to advance the research focus on human-centered artificial intelligence and reliable, trusted interfaces.

\begin{acks}
The support for this work is provided by DFKI and is partially funded by the German Ministry of Education and Research (BMBF) under project TeachTAM (Grant Number: 01IS17043) and project CAMELOT (Grant Number: 01IW20008), to which I am grateful for the opportunity. I would like to thank my doctoral advisor Dr. Michael Feld and my doctoral supervisor Prof. Dr. Antonio Kr{\"u}ger, for their constant support and constructive feedback.
\end{acks}

\bibliographystyle{ACM-Reference-Format}
\bibliography{sample-base}


\end{document}